# Observation of Terahertz Spin Hall Conductivity Spectrum in GaAs with Optical Spin Injection

Tomohiro Fujimoto*, Takayuki Kurihara, Yuta Murotani, Tomohiro Tamaya, Natsuki Kanda, Changsu Kim, Jun Yoshinobu, Hidefumi Akiyama, Takeo Kato, and Ryusuke Matsunaga*

*The Institute for Solid State Physics, The University of Tokyo, Kashiwa, Chiba 277-8581, Japan*
*e-mail: fujimoto@issp.u-tokyo.ac.jp, matsunaga@issp.u-tokyo.ac.jp

## Abstract

We report the first observation of the spin Hall conductivity spectrum in GaAs at room temperature. Our terahertz polarimetry with a precision of several µrads resolves the Faraday rotation of terahertz pulses arising from the inverse spin Hall effect of optically injected spin-polarized electrons. The obtained spin Hall conductivity spectrum exhibits an excellent quantitative agreement with theory, demonstrating a crossover in the dominant origin from impurity scattering in the DC regime to the intrinsic Berry-curvature mechanism in the terahertz regime. Our spectroscopic technique opens a new pathway to analyze anomalous transports related to spin, valley, or orbital degrees of freedom.

In the spin-orbit coupled systems, the current flow under a bias field is deflected in transverse directions dependent on the carrier spin, thus giving rise to the transverse spin current. The spin Hall effect (SHE) and its inverse process, the inverse spin Hall effect (ISHE), are key components in the conversion between charge and spin currents in spintronics [1–9]. The concept of spintronics, which utilizes spin as an additional degree of freedom for electrons, has been further extended to valleys in momentum space [10,11] and to orbital angular momentum [12,13]. The microscopic mechanisms of the SHE and ISHE, associated with the anomalous Hall effect (AHE) in magnets [14], have been intensively investigated in terms of the extrinsic impurity scattering [1-3] and dissipation-less intrinsic mechanism [4,5]. While most studies on the SHE and ISHE have focused on the quasi-static response under a DC bias field, the dynamics of the SHE and ISHE at timescales comparable to or faster than the spin relaxation are yet to be investigated. Quantum interference using near-infrared (NIR) femtosecond pulses has been demonstrated to enable ultrafast optical control of charge and spin currents [15]. Optically excited spin-polarized carriers in magnetic/nonmagnetic metal heterostructure thin films can be converted into ultrashort in-plane transverse currents; thus, they are attracting attention as broadband terahertz (THz) emitters [16]. However, studies on SHE and ISHE driven by high-speed electric fields are lacking. Even in a material where the extrinsic mechanisms dominate the DC transport, the AC Hall response can be solely driven by the intrinsic mechanism if the driving electric field is faster than the scattering rate, which typically lies in the THz frequency region. In line with the recent enthusiasm for ultrafast spintronics, such as antiferromagnets exploiting spin precessional motion in THz frequencies and control of magnetism using spin-orbit or spin-transfer torques [17-19], the frequency characteristics of spin-charge current conversion need to be clarified for the development of high-speed spintronic applications.

As the spintronic properties of materials can be sensitive to adjacent magnets [20], the noncontact optical injection of spins using circularly polarized light is particularly important. The conversion of spin-polarized photocarriers to Hall currents by the ISHE has been studied in the form of the light-induced AHE in the DC limit [21–25]. Polarization rotation of THz pulses can also be a probe for the ISHE to reveal its dynamical aspects and optically injected Berry curvature [26]. However, to the best of our knowledge, THz spectroscopy has been limited to the study of a semiconductor GaAs quantum well at cryogenic temperatures [27]. Although the phase shift of THz transients has been discussed in time domain [27], the phase shift in the limited time window is mostly determined by the response at the peak frequency of its spectral weight, which significantly degrades spectral resolution. Because the extrinsic contribution would sharply depend on frequency and the impurity scattering rate, the microscopic origin should be discussed from the spectral profile of optical conductivity. For this purpose, the full waveform of polarization-rotated THz pulse is required with a higher signal-to-noise ratio. Spin Hall current dynamics in bulk GaAs was also studied from the THz pulse emission [28,29]; however, spectral analysis was difficult owing to coexisting contributions from the surface and bulk, as well as the propagation effect. For a comprehensive understanding of this dynamical Hall conductivity, a quantitative analysis of its frequency characteristics, *i.e.* the spin Hall conductivity spectrum, is required, as studied for the anomalous Hall conductivity spectrum in magnets [30–35].

In this Letter, we have conducted NIR circularly polarized pump-THz probe experiments for bulk semiconductor GaAs to reveal the dynamical properties of spin transport. The anomalous Faraday rotation of the THz probe pulse depending on the pump helicity is clearly observed as a manifestation of the spin-to-charge current conversion, namely the ISHE. The signal shows double-exponential decays with time constants consistent with the spin relaxation of the valence and conduction bands. After the hole contribution relaxes, the spin Hall conductivity spectrum of the electrons is determined experimentally by suppressing the noise of polarization rotation angle to several μrad. Theoretical calculations of the sum of the intrinsic and extrinsic mechanisms quantitatively reproduces the experimentally observed spectrum quite well. This work clearly resolves the microscopic mechanisms of SHE from the frequency characteristics.

Figures 1(a) and 1(b) show schematics of the present pump-probe spectroscopy setup and experimental geometry. The circularly polarized NIR pump pulse excites the spin-polarized carriers in GaAs. According to the selection rule for interband transitions, left-handed circularly polarized (LCP) photons, denoted by $\sigma_+$, excite up- and down-spin electrons from the light-hole (LH) and heavy-hole (HH) bands with the angular momenta

of –(3/2)$\hbar$ and –(1/2)$\hbar$, respectively; the signs are the opposite for right-handed circularly polarized (RCP) photons. As the oscillator strength of transitions from the HH band to the conduction band is three times larger than that for the LH, the spin polarization ratio of excited electrons $P_s \equiv (N_\uparrow - N_\downarrow)/(N_\uparrow + N_\downarrow)$ is –0.5 [36], where $N_\uparrow$ and $N_\downarrow$ are the densities of up- and down-spin electrons. Subsequently, the carriers with spin polarization are driven by a THz electric field linearly polarized in the $x$-direction, thus yielding a net charge current $J_y$ in the $y$-direction owing to the ISHE.

The sample is undoped (001) GaAs grown by the molecular-beam epitaxy method with a thickness of 1.0 μm, sandwiched by $Al_{0.19}Ga_{0.81}As$ protective layers. All the experiments are performed at room temperature. To perform NIR pump-THz probe spectroscopy, we use a Yb:KGW laser amplifier system. Part of the output beam is converted into the NIR pump pulse at 1.46 or 1.55 eV using an optical parametric amplifier. Given that the bandgap of $Al_{0.19}Ga_{0.81}As$ is 1.63 eV, the NIR pump excites only the GaAs layer with a bandgap of 1.42 eV. The pump pulse duration is 200 fs, and the polarization is controlled using a quarter-wave plate to switch between LCP and RCP. The remnant of the laser output is compressed to 100 fs using the multiplate broadening scheme [37,38] and split into two beams for the generation and detection of the THz pulse in the form of time-domain spectroscopy. The polarization of THz probe is linearly aligned in the *x*-direction. After transmitting through the sample, the probe THz pulse is detected by the gate pulse in electro-optic sampling. A pair of wire-grid polarizers is inserted between the sample and the detection crystal to separately detect the *x*- and *y*-components of the THz probe pulse. The details of present experiments and analyses are described in the Supplemental Material [39].

The use of a thin sample allows the quantitative analysis of the response function through transmission measurements, maintaining the bulk nature. Two delay stages are used to control the delay times of the pump and probe pulses. We denote the time difference between the probe and gate pulses as $t_1$, and that between the pump and gate as $t_2$, as shown in Fig. 2(a). The upper panel of Fig. 2(b) shows $E_x(t_1)$, the probe THz pulse waveform transmitted through the sample without a pump. The lower panel shows a two-dimensional (2D) plot of $E_x$ as a function of $t_1$ and $t_2$ with the pump at 1.46 eV. The amplitude of $E_x$ decreases upon pumping, indicating that the transmission is suppressed by the photoexcited carriers. Figure 2(c) shows the photoinduced longitudinal conductivity spectrum $\Delta\sigma_{xx}(\omega)$ at $t_2 = 5.0$ ps with a pump fluence of 19.7 μJ cm$^{-2}$. Fitting by the Drude model gives the excited carrier density $N$ as shown in Fig. 2(d). We find that $N$ increases linearly with the fluence below 100 μJ cm$^{-2}$, thus indicating the linear absorption regime. Figure 2(e) shows the $t_2$ dependence of $N$. The decay time of $N$, which corresponds to the carrier recombination time, is longer than 1 ns and can be neglected in the timescale considered below.

Next, we discuss the dynamics of anomalous Faraday rotation signal. Figure 3(a) shows a 2D plot of the $y$ component of the transmitted THz pulse, $\Delta E_y(t_1, t_2)$, at a pump fluence of 18.3 μJ cm$^{-2}$ and a pump photon energy of 1.55 eV. To extract the signal depending on the pump helicity, we define $\Delta E_y(t_1, t_2)$ as half of the difference between the results for the LCP and RCP pumps: $\Delta E_y(t_1, t_2) \equiv (E_y^{\text{LCP}}(t_1, t_2) - E_y^{\text{RCP}}(t_1, t_2))/2$. Figure 3(a) shows that $\Delta E_y$ arises upon pumping and most of the signal decays quickly at $t_2 < 1$ ps. In addition, a small portion of the signal survives with a much longer lifetime at $t_2 > 1$ ps. Further, a slight phase shift is observed between the fast ($t_2 < 1$ ps) and slow ($t_2 > 1$ ps) signals, suggesting the different origin of the response in the fast and slow dynamics. For analysis, $t_1$ is fixed at 0.12 and 0.30 ps, the peak of fast and slow components, respectively. Their time evolutions after the pump are investigated by scanning along $t_2$. Figure 3(b) shows the quickly decaying signal as a function of $t_2$. The data are fitted using an exponential function with an offset, which yields a decay time of approximately 200 fs. By contrast, the slow decay signals are fitted by an exponential function with a time constant of 80 ps, as shown in Fig. 3(c). The decay times do not depend on the pump photon energy between 1.46 and 1.55 eV. Previously, the spin dynamics in bulk GaAs have been investigated using optical pump-optical probe spectroscopy, and the typical spin relaxation times of holes and electrons at room temperature were reported to be approximately 100 fs [49] and 75 ps [50], respectively. These spin relaxation times are in good agreement with the two decay times observed in the present experiment. Therefore, the fast and slowly decaying $\Delta E_y$ signals can be ascribed to the ISHE of optically-injected spin-polarized holes and electrons, respectively. The polarization rotation angle for the ISHE signal of electrons is so small as 100 µrad [39], which can be ascribed to the small spin-orbit coupling in the *s*-like conduction band. The upper and lower panels of Fig. 3(d) show the decay times of the fast (hole) and slow (electron) signals, respectively, as functions of the pump fluence. The decay times are almost independent of the pump fluence, thus indicating that the spin relaxation time does not depend on the carrier density in the measured region.

Note that, when the circularly polarized light temporally overlaps with the $x$-polarized THz field, a photocurrent in the $y$-direction can be generated owing to the anisotropic distribution of photoexcited carriers in momentum space, which was recently revealed in graphene [51] and a Dirac semimetal [52]. As this effect is well described by field-induced nonlinear current generation rather than a polarization rotation of the THz pulse, the resulting 2D map signal of $\Delta E_y(t_1, t_2)$ appears in a different way. In the present experimental condition this photocurrent effect is negligibly small, which will be presented elsewhere.

To quantitatively evaluate the anomalous Hall conductivity spectrum $\sigma_{yx}(\omega)$, we measure the transmitted THz signal at $t_2 = 5.0$ ps, where the faster ISHE signal of the holes almost completely vanishes and only the slower electron contribution remains at a nearly constant value. For a spectral analysis, we suppress the statistic errors to be several µrad and measure the whole waveform of $E_y(t_1)$ with a high signal-to-noise ratio [39]. $\sigma_{yx}(\omega)$ is obtained using the following equation.

$$\sigma_{yx}(\omega) = \frac{\Delta\sigma_{xx}(\omega)}{E_x^{\text{neq}}(\omega)/E_x^{\text{eq}}(\omega) - 1} \tilde{\theta}(\omega), \quad (1)$$

where $E_x^{\text{neq}}(\omega)$ and $E_x^{\text{eq}}(\omega)$ are the $E_x$ spectra with and without pumping, respectively, and $\tilde{\theta}(\omega)$ is the polarization rotation spectrum [32]. Furthermore, using the spin-polarization ratio $P_s$, the anomalous Hall conductivity can be converted to the spin Hall conductivity using the relation, $\sigma_{yx}^{\text{SH}}(\omega) = (P_s)^{-1}\sigma_{yx}(\omega)$. Here, $P_s$ can be fixed at –0.5 because the excitation process lies within the linear regime [Fig. 2(d)] and because the electron spin relaxation is negligible at $t_2 = 5.0$ ps owing to a much longer relaxation time [Fig. 3(d)]. Figures 4(a) and 4(b) show the real and imaginary parts of the spin Hall conductivity spectrum $\sigma_{yx}^{SH}(\omega)$, respectively, with the pump at 1.46 eV. The fluence is the same as the experiment for $\sigma_{xx}(\omega)$ in Fig. 2(c).

According to theoretical studies on the frequency dependence of the AHE and SHE in the conduction band of semiconductors [52,53], each contribution of the intrinsic and side-jump mechanisms can be represented by:

$$\sigma_{yx}^{SH,int}(\omega) = 2N\lambda \frac{e^2}{\hbar}, \tag{2}$$

$$\sigma_{yx}^{SH,sj}(\omega) = -4N\lambda \frac{e^2}{\hbar} \frac{1}{1 - i\omega\tau_{ex}}, \tag{3}$$

where $N$ is the electron density, $\lambda$ is the spin-orbit coupling constant, and $\tau_{ex}$ is the extrinsic scattering time [39]. In addition, the skew scattering contribution can be approximately given by:

$$\sigma_{yx}^{SH,skew}(\omega) \approx 2N\lambda \frac{e^2}{\hbar} \frac{E_B}{\hbar} \tau_{ex} \frac{1}{(1 - i\omega\tau_{ex})^2}, \tag{4}$$

where $E_B$ is the binding energy of the impurity potential [39]. The sum of Eqs. (2)-(4) has successfully explained the previous experimental results of SHE in the DC limit at room temperature [25]. For the bulk GaAs, the known parameters are: $\lambda$ = 5.3 Å$^2$ [54] and $E_B = m^*R_y/m_0\epsilon^2$ = 5.5 meV, where $R_y$ is the Rydberg constant, $m_0$ is the free electron mass, $m^*$ = 0.067$m_0$, and $\epsilon$ = 12.9 are the effective mass and permittivity, respectively [55,56]. From the Drude model fitting for $\sigma_{xx}(\omega)$ in Fig. 2(c), the electron density $N$ and relaxation time $\tau$ were obtained as $N$ = 1.3 × 10$^{17}$ cm$^{-3}$ and $\tau$ = 150 fs. The impurity scattering time $\tau_{ex}$ in Eqs. (3) and (4) is expected to be the same as or a bit longer than $\tau$. In Figs. 4(c) and 4(d), we plot $\sigma_{yx}^{SH}(\omega)$ as the sum of Eqs. (2)-(4), using $\tau_{ex}$ = 150, 250, and 350 fs. Note that, except for $\tau_{ex}$, any fitting parameters are not used in the calculation. The experimental results of $\sigma_{yx}^{SH}(\omega)$ in THz frequency are well reproduced by the calculations for any value of $\tau_{ex}$ between 150 and 350 fs. It is in contrast to the DC limit, which is sensitive to $\tau_{ex}$ because of the large influence of skew scattering. The result suggests that the SHE in THz frequency is less dependent on the scattering.

Using $\tau_{\text{ex}}$ = 250 fs, Figs. 4(e) and 4(f) show the real and imaginary parts of the calculated $\sigma_{yx}^{\text{SH}}(\omega)$, respectively, for each contribution of the intrinsic, side-jump, and skew scattering mechanisms. In the DC limit, the intrinsic contribution is canceled by half of the side-jump contribution such that the total spin Hall conductivity of the electrons is dominated by extrinsic impurity scattering, which is consistent with the previous static measurement of the SHE for $n$-doped bulk GaAs [6]. At sub-THz frequency, the spin Hall conductivity decreases because the impurity scattering is suppressed when the electric field alternates faster than the scattering rate. As the frequency increases beyond 1 THz, however, the real-part spin Hall conductivity recovers to the value comparable to, or even larger than, that in the DC limit owing to the dominant contribution of intrinsic Berry curvature mechanism, which is nondissipative and independent of frequency. Although the imaginary part of the side-jump effect is still considerable at approximately 1 THz, it would be suppressed in higher frequency [39]. The dissipation is rather dominated by the longitudinal current. Owing to the sharp Drude response in Fig. 2(c), the longitudinal conductivity $\Delta\sigma_{xx}$ is also suppressed and becomes less dissipative at several THz. Therefore, the larger spin Hall angle $\theta_{\text{SH}} = \sigma_{yx}^{\text{SH}}/\sigma_{xx}$ can be expected with less dissipative nature at several THz, thus implying an efficient spin-to-charge current conversion using semiconductors.

In conclusion, the spin Hall conductivity spectrum in GaAs at room temperature was successfully observed using our highly precise THz polarimetry. The excellent agreement between the experiment and theory in the representative material will stimulate further exploration of the spin Hall conductivity spectrum of various materials to reveal their microscopic origin, as spin-polarized carriers can be injected by light even in heavy metals [57]. The extension to a faster electric field up to the multi-THz frequency [38,58,59] is also promising for materials with a faster scattering time, such as transition metals with a much larger SHE [60]. The nonlinearity expected for an intense THz electric field and its effect on scattering [25,61,62] are also highly intriguing. This study opens a new avenue for ultrafast noncontact detection schemes for the anomalous transport related to spin, valley, and orbital degrees of freedom.

**Acknowledgements**

This work was supported by JST PRESTO (Grant Nos. JPMJPR20LA and JPMJPR2006), JST CREST (Grant No. JPMJCR20R4). R.M. acknowledges partial support from MEXT Quantum Leap Flagship Program (MEXT Q-LEAP, Grant No. JPMXS0118068681). T. Kurihara acknowledges the support from JSPS KAKENHI (Grant No. JP20K22478). T. T. acknowledges the support from JST PRESTO (Grant No. JPMJPR2107). NIR transmission measurements were performed using the facility of Materials Design and Characterization Laboratory in The Institute for Solid State Physics, The University of Tokyo. R.M., T. Kurihara, T.T., and T. Kato conceived the project. C.K., T.F., and H.A. fabricated the sample. T.F., Y.M., and N.K. developed the pump-probe spectroscopy system with the help of T. Kurihara, J.Y., and R.M. T.F. performed the pump-probe experiment and analyzed the data with Y.M. and T. Kurihara. All the authors discussed the results. T.F. and R.M. wrote the manuscript with substantial feedbacks from Y.M. and all the coauthors.

**References**

[1]     M. D'yakonov and V. Perel, Possibility of Orienting Electron Spins with Current, JETP Lett. **13**, 467 (1971).

[2]     J. E. Hirsch, Spin Hall Effect, Phys. Rev. Lett. **83**, 1834 (1999).

[3]     S. Zhang, Spin Hall Effect in the Presence of Spin Diffusion, Phys. Rev. Lett. **85**, 393 (2000).

[4]     S. Murakami, N. Nagaosa, and S.-C. Zhang, Dissipationless Quantum Spin Current at Room Temperature, Science **301**, 1348 (2003).

[5]     J. Sinova, D. Culcer, Q. Niu, N. A. Sinitsyn, T. Jungwirth, and A. H. MacDonald, Universal Intrinsic Spin Hall Effect, Phys. Rev. Lett. **92**, 126603 (2004).

[6]     Y. K. Kato, R. C. Myers, A. C. Gossard, and D. D. Awschalom, Observation of the Spin Hall Effect in Semiconductors, Science **306**, 1910 (2004).

[7]     J. Wunderlich, B. Kaestner, J. Sinova, and T. Jungwirth, Experimental Observation of the Spin-Hall Effect in a Two-Dimensional Spin-Orbit Coupled Semiconductor System, Phys. Rev. Lett. **94**, 047204 (2005).

[8]     E. Saitoh, M. Ueda, H. Miyajima, and G. Tatara, Conversion of spin current into

charge current at room temperature: Inverse spin-Hall effect, Appl. Phys. Lett. **88**, 182509 (2006).

[9] J. Sinova, S. O. Valenzuela, J. Wunderlich, C. H. Back, and T. Jungwirth, Spin Hall effects, Rev. Mod. Phys. **87**, 1213 (2015).

[10] A. Rycerz, J. Tworzydło, and C. W. J. Beenakker, Valley filter and valley valve in graphene, Nature Phys. **3**, 172 (2007).

[11] K. F. Mak, K. L. McGill, J. Park, and P. L. McEuen, The valley Hall effect in $MoS_2$ transistors, Science **344**, 1489 (2014).

[12] B. A. Bernevig, T. L. Hughes, and S.-C. Zhang, Orbitronics: The Intrinsic Orbital Current in *p*-Doped Silicon, Phys. Rev. Lett. **95**, 066601 (2005).

[13] D. Go, D. Jo, H.-W. Lee, M. Kläui, and Y. Mokrousov, Orbitronics: Orbital currents in solids, EPL (Europhys. Lett.) **135**, 37001 (2021).

[14] N. Nagaosa, J. Sinova, S. Onoda, A. H. MacDonald, and N. P. Ong, Anomalous Hall effect, Rev. Mod. Phys. **82**, 1539 (2010).

[15] H. Zhao, E. J. Loren, H. M. van Driel, and A. L. Smirl, Coherence Control of Hall Charge and Spin Currents, Phys. Rev. Lett. **96**, 246601 (2006).

[16] T. Seifert *et al.*, Efficient metallic spintronic emitters of ultrabroadband terahertz radiation, Nature Photon. **10**, 483 (2016).

[17] V. Baltz, A. Manchon, M. Tsoi, T. Moriyama, T. Ono, and Y. Tserkovnyak, Antiferromagnetic spintronics, Rev. Mod. Phys. **90**, 015005 (2018).

[18] L. Šmejkal, Y. Mokrousov, B. Yan, and A. H. MacDonald, Topological antiferromagnetic spintronics, Nature Phys. **14**, 242 (2018).

[19] J. Han, R. Cheng, L. Liu, H. Ohno, and S. Fukami, Coherent antiferromagnetic spintronics, Nature Mater. **22**, 684 (2023).

[20] C. Stamm, C. Murer, M. Berritta, J. Feng, M. Gabureac, P. M. Oppeneer, and P. Gambardella, Magneto-Optical Detection of the Spin Hall Effect in Pt and W Thin Films, Phys. Rev. Lett. **119**, 087203 (2017).

[21] A. A. Bakun, B. P. Zakharchenya, A. A. Rogachev, M. N. Tkachuk, and V. G. Fleisher, Observation of a surface photocurrent caused by optical orientation of electrons in a semiconductor, JETP Lett. **40**, 1293 (1984).

[22] M. I. Miah, Observation of the anomalous Hall effect in GaAs, J. Phys. D: Appl. Phys. **40**, 1659 (2007).

[23]     C. M. Yin *et al.*, Observation of the photoinduced anomalous Hall effect in GaN-based heterostructures, Appl. Phys. Lett. **98**, 122104 (2011).

[24]     J. L. Yu, Y. H. Chen, C. Y. Jiang, Y. Liu, H. Ma, and L. P. Zhu, Observation of the photoinduced anomalous Hall effect spectra in insulating InGaAs/AlGaAs quantum wells at room temperature, Appl. Phys. Lett. **100**, 142109 (2012).

[25]     N. Okamoto, H. Kurebayashi, T. Trypiniotis, I. Farrer, D. A. Ritchie, E. Saitoh, J. Sinova, J. Mašek, T. Jungwirth, and C. H.W. Barnes, Electric control of the spin Hall effect by intervalley transitions, Nature Mater. **13**, 932 (2014).

[26]     K. S. Virk and J. E. Sipe, Optical Injection and Terahertz Detection of the Macroscopic Berry Curvature, Phys. Rev. Lett. **107**, 120403 (2011).

[27]     S. Priyadarshi, K. Pierz, and M. Bieler, Detection of the Anomalous Velocity with Subpicosecond Time Resolution in Semiconductor Nanostructures, Phys. Rev. Lett. **115**, 257401 (2015).

[28]     C. B. Schmidt, S. Priyadarshi, S. A. Tarasenko, and M. Bieler, Ultrafast magneto-photocurrents in GaAs: Separation of surface and bulk contributions, Appl. Phys. Lett. **106**, 142108 (2015).

[29]     C. B. Schmidt, S. Priyadarshi, and M. Bieler, Sub-picosecond temporal resolution of anomalous Hall currents in GaAs, Sci. Rep. **7**, 11241 (2017).

[30]     Z. Fang, N. Nagaosa, K. S. Takahashi, A. Asamitsu, R. Mathieu, T. Ogasawara, H. Yamada, M. Kawasaki, Y. Tokura, and K. Terakura, The Anomalous Hall Effect and Magnetic Monopoles in Momentum Space, Science **302**, 92 (2003).

[31]     R. Shimano, Y. Ikebe, K. S. Takahashi, M. Kawasaki, N. Nagaosa, and Y. Tokura, Terahertz Faraday rotation induced by an anomalous Hall effect in the itinerant ferromagnet $SrRuO_3$, EPL (Europhys. Lett.) **95**, 17002 (2011).

[32]     T. Matsuda, N. Kanda, T. Higo, N. P. Armitage, S. Nakatsuji, and R. Matsunaga, Room-temperature terahertz anomalous Hall effect in Weyl antiferromagnet $Mn_3Sn$ thin films, Nature Commun. **11**, 909 (2020).

[33]     Y. Okamura *et al.*, Giant magneto-optical responses in magnetic Weyl semimetal $Co_3Sn_2S_2$, Nature Commun. **11**, 4619 (2020).

[34]     T. S. Seifert *et al.*, Frequency-Independent Terahertz Anomalous Hall Effect in $DyCo_5$, $Co_{32}Fe_{68}$, and $Gd_{27}Fe_{73}$ Thin Films from DC to 40 THz, Adv. Mater. **33**, 2007398 (2021).

[35] T. Matsuda, T. Higo, T. Koretsune, N. Kanda, Yoshua Hirai, Hanyi Peng, Takumi Matsuo, Naotaka Yoshikawa, Ryo Shimano, S. Nakatsuji, and R. Matsunaga, Ultrafast Dynamics of Intrinsic Anomalous Hall Effect in the Topological Antiferromagnet $Mn_3Sn$, Phys. Rev. Lett. **130**, 126302 (2023).

[36] I. Žutić, J. Fabian, and S. Das Sarma, Spintronics: Fundamentals and applications, Rev. Mod. Phys. **76**, 323 (2004).

[37] C.-H. Lu, Y.-J. Tsou, H.-Y. Chen, B.-H. Chen, Y.-C. Cheng, S.-D. Yang, M.-C. Chen, C.-C. Hsu, and A. H. Kung, Generation of intense supercontinuum in condensed media, Optica **1**, 400 (2014).

[38] N. Kanda, N. Ishii, J. Itatani, and R. Matsunaga, Optical parametric amplification of phase-stable terahertz-to-mid-infrared pulses studied in the time domain, Opt. Express **29**, 3479 (2021).

[39] See Supplemental Material for sample preparation, experimental methods, and analysis, which includes Refs. [40–48].

[40] F. Sekiguchi, T. Mochizuki, C. Kim, H. Akiyama, L. N. Pfeiffer, K. W. West, and R. Shimano, Anomalous Metal Phase Emergent on the Verge of an Exciton Mott Transition, Phys. Rev. Lett. **118**, 067401 (2017).

[41] N. Kanda, K. Konishi, and M. Kuwata-Gonokami, Terahertz wave polarization rotation with double layered metal grating of complimentary chiral patterns, Opt. Express **15**, 11117 (2007).

[42] R. A. Kaindl, D. Hägele, M. A. Carnahan, and D. S. Chemla, Transient terahertz spectroscopy of excitons and unbound carriers in quasi-two-dimensional electron-hole gases, Phys. Rev. B **79**, 045320 (2009).

[43] C. Xiao, B. Xiong, and F. Xue, Boltzmann approach to spin–orbit-induced transport in effective quantum theories, J. Phys.: Condens. Matter **30**, 415002 (2018).

[44] N. A. Sinitsyn, A. H. MacDonald, T. Jungwirth, V. K. Dugaev, and J. Sinova, Anomalous Hall effect in a two-dimensional Dirac band: The link between the Kubo-Streda formula and the semiclassical Boltzmann equation approach, Phys. Rev. B **75**, 045315 (2007).

[45] E. S. Garlid, Q. O. Hu, M. K. Chan, C. J. Palmstrøm, and P. A. Crowell, Electrical Measurement of the Direct Spin Hall Effect in Fe/$In_xGa_{1-x}As$ Heterostructures, Phys. Rev. Lett. **105**, 156602 (2010).

[46] H. Němec, F. Kadlec and P. Kužel, Methodology of an optical pump-terahertz probe experiment: An analytical frequency-domain approach, J. Chem. Phys. **117**, 8454 (2002).

[47] M. C. Beard, G. M. Turner and C. A. Schmuttenmaer, Transient photoconductivity in GaAs as measured by time-resolved terahertz spectroscopy, Phys. Rev. B **62**, 15764 (2000).

[48] M. Sotoodeh, A. H. Khalid and A. A. Rezazadeh, Empirical low-field mobility model for III–V compounds applicable in device simulation codes, J. Appl. Phys. 87, 2890 (2000).

[49] D. J. Hilton and C. L. Tang, Optical Orientation and Femtosecond Relaxation of Spin-Polarized Holes in GaAs, Phys. Rev. Lett. **89**, 146601 (2002).

[50] S. Oertel, J. Hübner, and M. Oestreich, High temperature electron spin relaxation in bulk GaAs, Appl. Phys. Lett. **93**, 132112 (2008).

[51] S. A. Sato *et al.*, Microscopic theory for the light-induced anomalous Hall effect in graphene, Phys. Rev. B **99**, 214302 (2019).

[52] Y. Murotani, N. Kanda, T. Fujimoto, T. Matsuda, M. Goyal, J. Yoshinobu, Y. Kobayashi, T. Oka, S. Stemmer, and R. Matsunaga, Disentangling the Competing Mechanisms of Light-Induced Anomalous Hall Conductivity in Three-Dimensional Dirac Semimetal, Phys. Rev. Lett. **131**, 096901 (2023).

[53] P. Nozières and C. Lewiner, A simple theory of the anomalous hall effect in semiconductors, J. Phys. France **34**, 901 (1973).

[54] H.-A. Engel, B. I. Halperin, and E. I. Rashba, Theory of Spin Hall Conductivity in *n*-Doped GaAs, Phys. Rev. Lett. **95**, 166605 (2005).

[55] J. S. Blakemore, Semiconducting and other major properties of gallium arsenide, J. Appl. Phys. **53**, R123 (1982).

[56] M. Cardona and Y. Y. Peter, *Fundamentals of semiconductors* (Springer, 2005), Vol. 619.

[57] G.-M. Choi, J. H. Oh, D.-K. Lee, S.-W. Lee, K. W. Kim, M. Lim, B.-C. Min, K.-J. Lee, and H.-W. Lee, Optical spin-orbit torque in heavy metal-ferromagnet heterostructures, Nat. Commun. **11**, 1482 (2020)

[58] N. Kanda, Y. Murotani, T. Matsuda, M. Goyal, S. Salmani-Rezaie, J. Yoshinobu, S. Stemmer, and R. Matsunaga, Tracking Ultrafast Change of Multiterahertz Broadband

Response Functions in a Photoexcited Dirac Semimetal $Cd_3As_2$ Thin Film, Nano Lett. **22**, 2358 (2022).

[59]     Y. Murotani, N. Kanda, T. N. Ikeda, T. Matsuda, M. Goyal, J. Yoshinobu, Y. Kobayashi, S. Stemmer, and R. Matsunaga, Stimulated Rayleigh Scattering Enhanced by a Longitudinal Plasma Mode in a Periodically Driven Dirac Semimetal $Cd_3As_2$, Phys. Rev. Lett. **129**, 207402 (2022).

[60]     T. Kimura, Y. Otani, T. Sato, S. Takahashi, and S. Maekawa, Room-Temperature Reversible Spin Hall Effect, Phys. Rev. Lett. **98**, 156601 (2007).

[61]     H. Hirori, K. Shinokita, M. Shirai, S. Tani, Y. Kadoya, and K. Tanaka, Extraordinary carrier multiplication gated by a picosecond electric field pulse, Nature Commun. **2**, 594 (2011).

[62]     W. Kuehn, P. Gaal, K. Reimann, M. Woerner, T. Elsaesser, and R. Hey, Coherent Ballistic Motion of Electrons in a Periodic Potential, Phys. Rev. Lett. **104**, 146602 (2010).

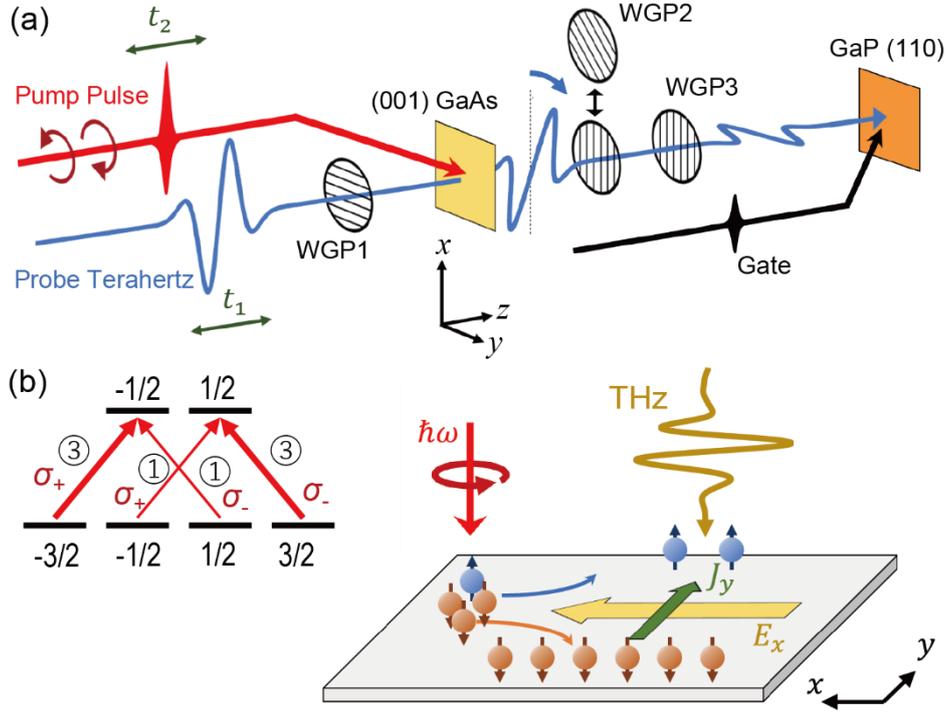

FIG. 1. (a) Schematic of the optical system. A circularly polarized pump pulse excites the sample, and the Faraday rotation of the THz probe pulse is detected by electro-optic sampling with a (110) GaP crystal. Both pulses are normally incident on the sample. (b) Schematic of the experimental configuration in the sample. $\sigma_+$ denotes the LCP pump, exciting spin-polarized electrons with $N_\downarrow/N_\uparrow=3$. When $x$-polarized THz field $E_x$ is applied, the net charge current $J_y$ is generated by the ISHE. Inset shows the selection rule of the interband transitions by the circularly polarized pump.

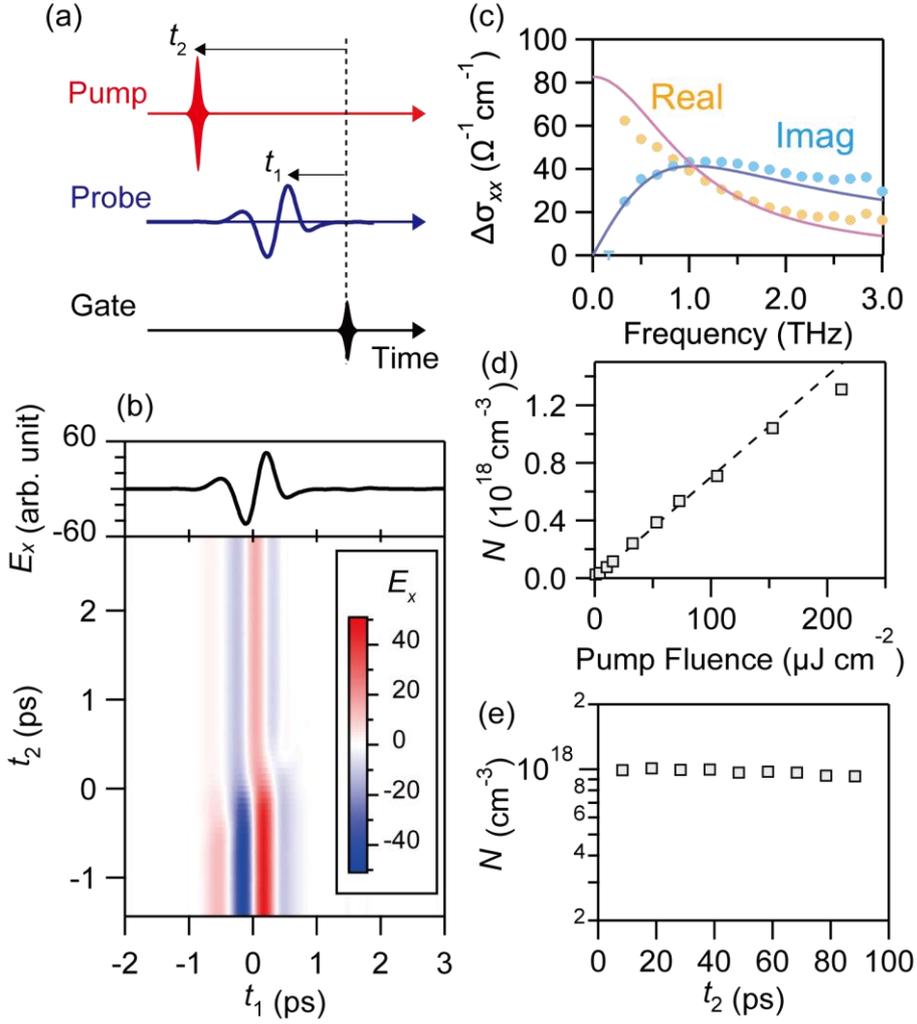

FIG. 2. (a) Schematic of relation between the pump, probe, and gate pulses. (b) (upper) THz electric field waveform $E_x(t_1)$ transmitting through the sample without the pump. (lower) 2D plot of the THz electric field $E_x(t_1, t_2)$ across the pump pulse irradiation around $t_2 \sim 0$ ps with a pump fluence of 109 μJ cm$^{-2}$ and a pump photon energy of 1.46 eV. (c) Solid circles show light-induced longitudinal conductivity spectrum $\Delta\sigma_{xx}(\omega)$ at $t_2 = 5.0$ ps with a pump fluence of 19.7 μJ cm$^{-2}$, whereas the solid curves show the Drude model fitting. Error bars represent statistical errors. (d),(e) Carrier density estimated from the Drude fitting as a function of the pump fluence at $t_2 = 5.0$ ps, and the pump delay $t_2$ with a pump fluence of 179 μJ cm$^{-2}$, respectively.

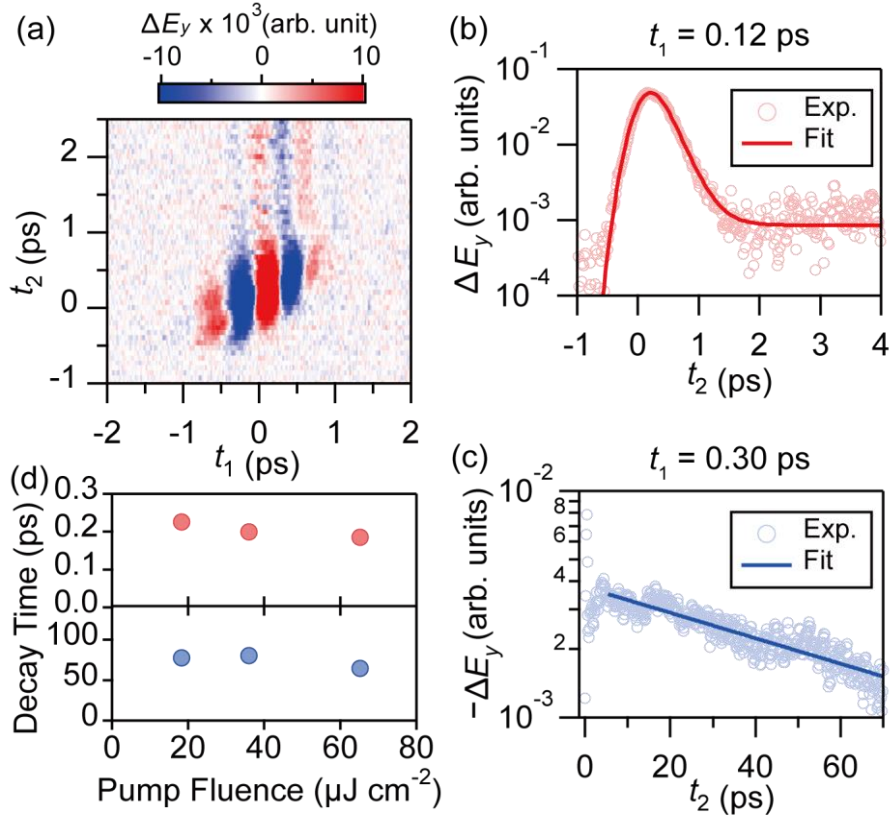

FIG. 3. (a) 2D plot of the light-induced $y$-polarized electric field $\Delta E_y$ as a function of $t_1$ and $t_2$ with a pump fluence of 18.3 µJ cm$^{-2}$ and a pump photon energy is 1.55 eV. (b),(c) Decay dynamics of $\Delta E_y$ for $t_1$=0.12 and 0.30 ps, respectively; the solid curves show the fitting results. (d) Upper and lower panels show the decay times as a function of the pump fluence for the data in (b) and (c), respectively.

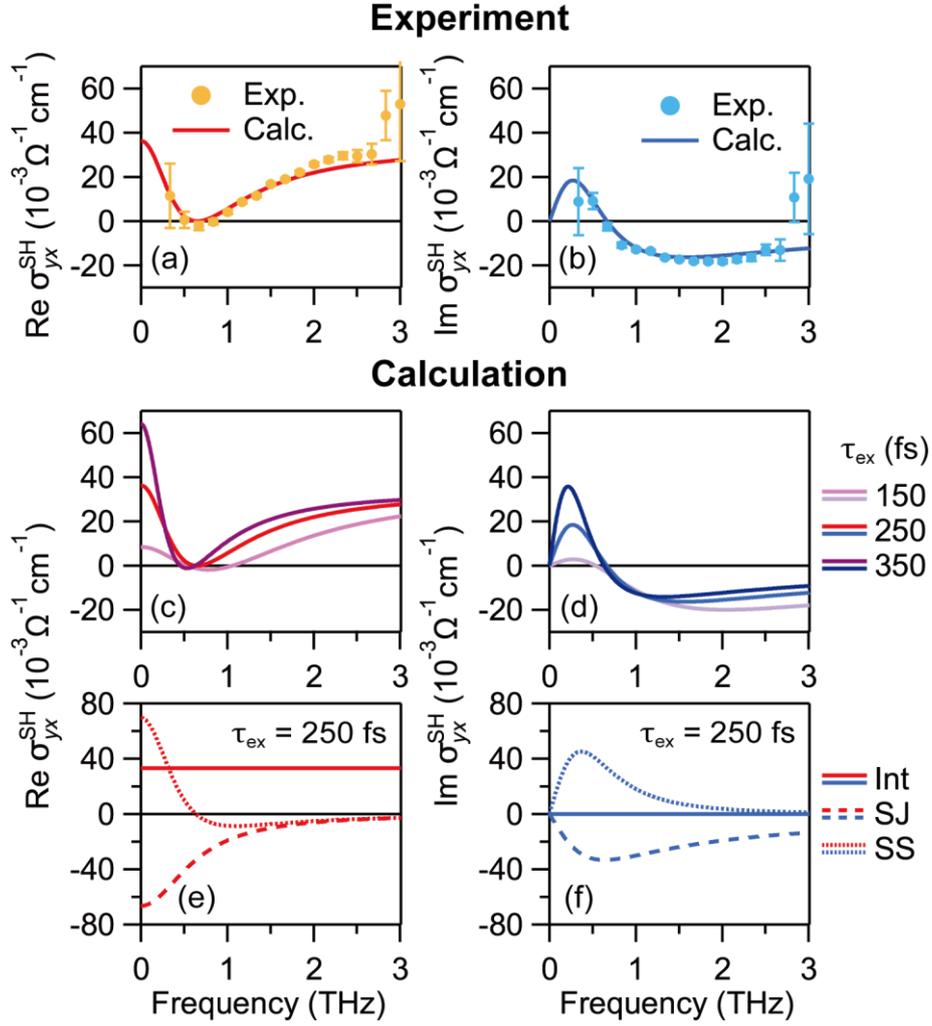

FIG. 4. (a),(b) Solid circles shows the experimental results of real and imaginary parts of the spin Hall conductivity spectrum $\sigma_{yx}^{SH}(\omega)$, respectively. Those error bars represent statistical errors. Solid curves show the theoretical curves for $\tau_{ex} = 250$ fs, using Eqs. (2)-(4). The pump photon energy is 1.46 eV and the fluence is 19.7 μJ cm$^{-2}$. (c),(d) Solid curves show the theoretical curves for various $\tau_{ex}$. (e),(f) Each microscopic contribution of the spin Hall conductivity for $\tau_{ex}$=250 fs. Int, SJ, and SS stand for the intrinsic, side-jump, and skew scattering mechanisms, respectively.